\documentclass[twocolumn]{elsart5p}
\usepackage{graphics}
\usepackage{graphicx}
\usepackage{dcolumn} 
\usepackage{bm}
\usepackage{epsfig}
\begin{document}
\begin{frontmatter}
\title{Origin of Strong Coupling in Lithium under Pressure}
\author[ucd]{Deepa Kasinathan,}
\author[ifw]{ K. Koepernik,}
\author[ucd,prag]{J. Kune\v{s},}
\author[mpi]{H. Rosner,}
\author[ucd]{W. E. Pickett \corauthref{cor}}
\corauth[cor]{Corresponding author.}
\ead{pickett@physics.ucdavis.edu}

\address[ucd]{Department of Physics, University of California Davis,
  Davis, CA 95616}
\address[ifw]{IFW Dresden, P. O. Box 270116, D-01171 Dresden, Germany}
\address[prag]{Institute of Physics, ASCR, Cukrovarnick\'a 10,
  162 53 Praha 6, Czech Republic}
\address[mpi]{Max-Planck-Institut f\"ur Chemische Physik
  fester Stoffe Dresden, Germany}

\begin{abstract}
In an attempt to provide a clearer understanding of the impressive
increase in T$_c$ under pressure in elemental Li,
linear response calculation of the phonon dispersion curves, electron-phonon
matrix elements, phonon linewidths and mode $\lambda$'s have been carried 
out on a finer mesh (24$^3$ in the Brillouin zone) than done previously,
for the volume corresponding to 20 GPa
pressure.  The result illustrates the great need for a 
fine mesh (even finer than this)
for converged results of $\lambda$ and the spectral function $\alpha^2 F$.
Although the initial pressure-induced transverse 
${\cal T}_1$ phonon instability 
(in harmonic approximation) near the
symmetry point K has dominated attention, the current results show
that the high value of T$_c$ gets strong
contributions from elsewhere in the zone, particularly from the 
longitudinal mode along (100). 
\end{abstract}
\end{frontmatter}                                                                                                                                                             
\maketitle
                                                                                                                                                             
\section{Introduction}
Elemental lithium, with superconducting T$_c \sim$ 15 K 
around\cite{shimizu,schilling,struzhkin} 30 GPa and up to 
20 K in strained samples\cite{shimizu} at 50 GPa, joins MgB$_2$ (T$_c$=40 K), metallic Y
(T$_c$=20 K)\cite{hamlin}, and PuCoGa$_5$ (T$_c$=19 K)\cite{sarrao} 
as the big surprises in critical
temperature in this century.  Transformation from a simple nearly-free-electron
metal not superconducting\cite{ucb} above 100 $\mu$K to a strongly coupled
electron-phonon superconductor at 20-50 GPa has been explained, at least
semiquantitatively, in recent work by Profeta {\it et al.}\cite{profeta} and by
Kasinathan {\it et al.}\cite{deepa}.
The evolution of T$_{c}$ has been linked to the
increase in electron-phonon coupling of specific phonon branches
along intersections of Kohn anomaly 
surfaces with pressure. The coupling strength is quite strongly dependent 
on the phonon wave vector $Q$ and phonon polarization. In particular, the $<1{\bar 1}0>$ polarized
phonon along the (110) $\Gamma$-K  direction softens with increasing 
pressure and finally becomes unstable. 

The nesting function $\xi(Q)$ which describes the phase 
space for the electron-hole scattering processes across the Fermi surface has
fine structure\cite{deepa} requiring a very dense mesh to obtain accurate values where its
amplitude is large.  Since the electron-phonon (EP) coupling involves these
same scattering events weighted by EP matrix elements that normally vary
smoothly with Q, the fine structure in $\xi(Q)$ implies corresponding fine
structure in EP coupling (as contained for example in the EP contribution
to the phonon linewidths $\gamma_{Q\nu}$).  Such fine structure makes the
sampling of the Brillouin zone an issue if accurate numerical results for
$\alpha^2 F$ are desired.
Our previous linear response phonon calculations \cite{deepa}
were performed on 12$^3$ mesh (72 $Q$ points in the irreducible zone). 
In the present work, we
focus on the 20 GPa case (fcc lattice constant a$_{0}$ = 6.80 bohr) 
where strong
phonon softening arises but structural instability has not yet occurred. The phonon
energies and the Eliashberg spectral function constants have been presently
calculated on a much finer 24$^3$ mesh (413 $Q$ points in the 
irreducible zone).  Careful examination
of the results lead to a revised picture of what coupling is important for T$_c$
(as opposed to what is important in increasing $\lambda$, which is a far
simpler question). \\

\section{Computational Details}

The electronic wave functions are calculated using a full potential linear 
muffin tin orbital (FP LMTO) method as implemented by Savrasov \cite{Sav}.
We have used the Vosko-Wilk-Nusair local exchange-correlation \cite{vwn}.
The phonon
energies and the coupling constants have been calculated using linear response
theory \cite{Sav}. Phonons were calculated for a dense 24$^3$ mesh 
corresponding to 413 inequivalent $Q$ points in the irreducible zone. 
The electronic BZ zone integration was performed with a 40$^3$ k point grid. \\

\section{Dispersion Throughout the Zone}

\begin{figure}
\vspace*{1cm}\rotatebox{-00}{\resizebox{8.5cm}{5.5cm}{\includegraphics{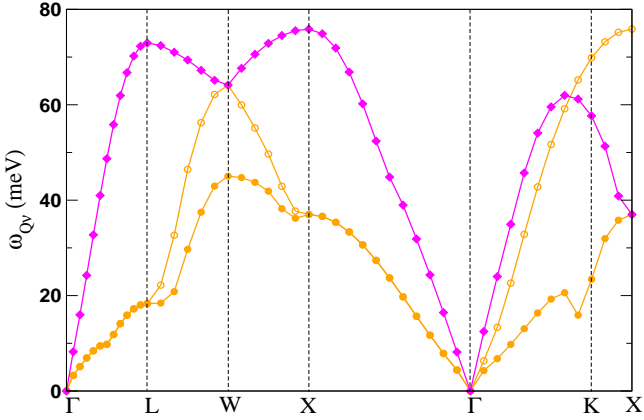}}}
\caption{(color online)
Calculated phonon spectrum for fcc Li at 20 GPa along high
symmetry directions.
The solid diamonds denote the longitudinal mode ${\cal L}$, and where
the transverse branches are non-degenerate, solid circles denote
the transverse mode ${\cal T}_1$ and open circles denote the transverse
mode ${\cal T}_2$.  The mode symmetry has been used to connect branches
across crossings ($\Gamma$-K) although this is not possible off symmetry
lines where branches do not cross. }
\label{phonons}
\end{figure}

The calculated phonon dispersion curves are displayed 
in Fig. \ref{phonons}. The only noticeable softening is seen
midway along the $\Gamma$-L direction (transverse ${\cal T}$ branches),
for the $<1{\bar 1}0>$ polarized ${\cal T}_1$ branch  
along the $\Gamma$-K direction emphasized previously,\cite{deepa}
and possibly softening in the ${\cal T}$ branches around the X point. 
We had identified previously that the $Q$ = $(\frac{2}{3},\frac{2}{3},0)
\frac{2\pi}{a}$ ${\cal T}_1$ phonon
(near K) causes large band shifts with atomic displacement
($\delta\varepsilon_k/\delta u_Q \approx$
5 eV/\AA) near the FS necks, reflecting strong EP coupling. 
 
\begin{figure}[tb]
\vspace*{1cm}\rotatebox{-00}{\resizebox{8.5cm}{5.5cm}{\includegraphics{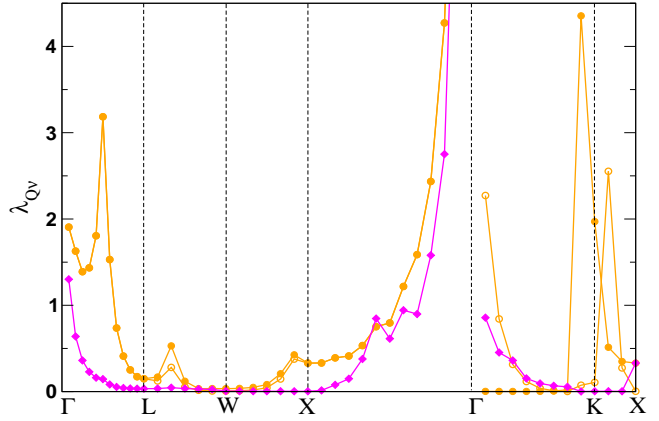}}}
\caption{(color online)
Calculated mode $\lambda_{Q\nu}$ values, following the notation of Fig.
\ref{phonons}.  The landscape is dominated by peaks in the transverse
branches near K, a longitudinal peak along $\Gamma$-L, and an increase
around Q$\sim$zero where phase space is limited.  Here $\lambda_{Q\nu}$ 
is normalized such that the total $\lambda$ requires a sum over the
three branches (rather than an average).
}
\label{modelambda}
\end{figure}

The corresponding `mode $\lambda$' values $\lambda_{Q,\nu}$ 
are presented in Fig. \ref{modelambda}. The $\lambda_{Q,\nu}$ values
for the ${\cal T}_1$ branch near K, and also near the midpoint of 
the $\Gamma$-L line are larger compared to other areas in the BZ, as
expected from the phonon kinks.  In addition, we 
notice that the $\lambda_{Q,\nu}$ values for all the three branches are
large all along $\Gamma$-X; this coupling is not immediately obvious  
from the phonon dispersion
in Fig. \ref{phonons} because it depends only weakly on Q. \\

\begin{figure}[tb]
\rotatebox{-00}{\resizebox{8.5cm}{5.5cm}{\includegraphics{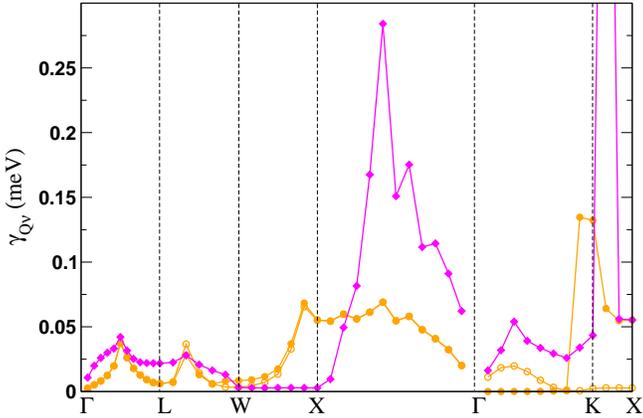}}}
\caption{(color online)
Calculated linewidths $\gamma_{Q\nu}$, following the notation of Fig.
\ref{phonons}.  The landscape here is dominated by all modes along
$\Gamma$-X, and the region near K.
}
\label{linewidth}
\end{figure}

In Fig. \ref{linewidth}, we display the corresponding linewidths 
$\gamma_{Q\nu}$, which are given by\cite{allen}
\begin{eqnarray}
\gamma_{Q\nu} = 2\pi \omega_{Q\nu} \sum_k |M_{k,k+Q}^{[\nu]}|^2
  \delta(\varepsilon_k)\delta(\varepsilon_{k+Q}).
\end{eqnarray}
While $\gamma_{Q\nu}$ includes the same Q-specific Fermi surface average 
of the EP matrix elements with respect to the available phase space
for scattering through wavevector Q as does $\lambda_{Q\nu}$, it is
proportional to $\omega_{Q\nu}$ rather than inversely proportional.
As a result, $\gamma_{Q\nu}$ is a much better indicator of the 
importance of the coupling for T$_c$.  Coupling at high frequency is
much more important than at low frequency for T$_c$, whereas the 
opposite is true for $\lambda$.  Thus it is not surprising that the
linewidth `dispersion' in Fig. \ref{linewidth} provides a different
viewpoint than does the $\lambda_{Q\nu}$ dispersion.  There is still the
large contribution from the region around K where there is an
incipient instability, but it is very limited in Q and very sharply 
structured, not being resolved even by the Q mesh we have used.  The really
impressive region however is the ${\cal L}$ branch along (100) directions,
which dominates the landscape (at least along symmetry lines).
There are also important contributions from the ${\cal L}$ branch 
along the $\Gamma$-K lines and from the ${\cal T}$ branches along
(100).  Even the small peak in the ${\cal T}$ branches midway
between $\Gamma$ and L (where the kink in the ${\cal T}$ branch occurs) is
small compared to the contribution from the ${\cal L}$ branch 
along the same line. Tse {\it et al.} present $\gamma_Q$ at V/V$_{0}$ = 
0.45 GPa ($\approx$ 50 GPa) which have some features \cite{tse} in common 
with Fig. \ref{linewidth}, but cannot be compared in detail because of the
different volumes. 
 
\section{Spectral Functions and T$_c$}
\begin{figure}[tb]
\vspace*{1cm}\rotatebox{-00}{\resizebox{8.5cm}{5.5cm}{\includegraphics{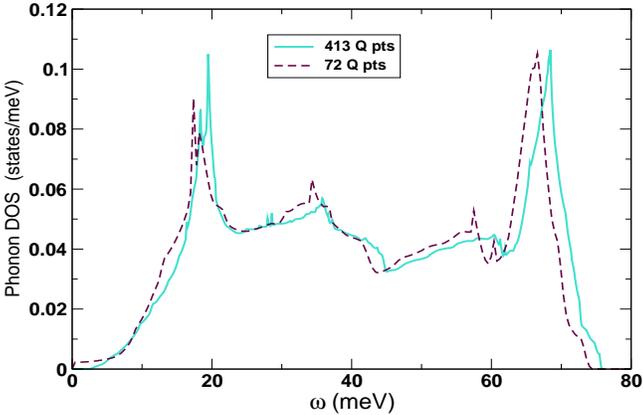}}}
\caption{(color online)
Phonon density of states calculated from the 12$^3$ Q grid (72 points) and
the 24$^3$ grid (413 points), illustrating that the representation of the
phonon spectrum changes little.  The substantial changes in $\alpha^2(\omega)$
shown in Fig. \ref{spectralfncomp} are discussed in the text.
}
\label{phonondoscomp}
\end{figure}

\begin{figure}[tb]
\rotatebox{-00}{\resizebox{8.5cm}{5.5cm}{\includegraphics{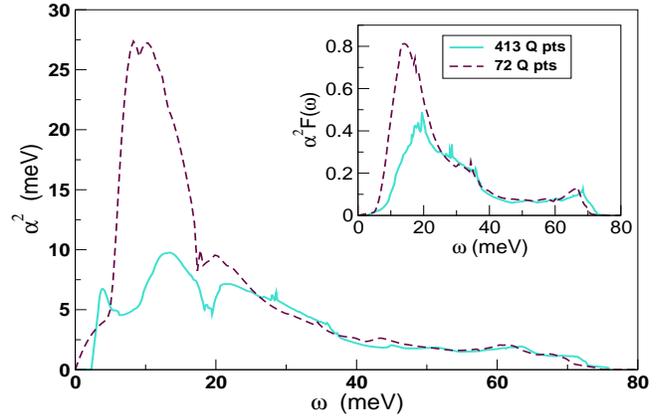}}}
\caption{(color online)
Comparison of the Eliashberg
spectral function $\alpha^{2}F(\omega)$ (inset) and the frequency-resolved 
coupling strength $\alpha^{2}(\omega)$ for fcc Li at 20 GPa, for both
Q point meshes.
}
\label{spectralfncomp}
\end{figure}

The phonon frequency distribution $F(\omega)$ is shown 
in Fig. \ref{phonondoscomp}
where it is clear that the coarser Q mesh had given a good picture
of the density of states.  This insensitivity does not persist when
coupling is included.
The Eliashberg spectral function  $\alpha^{2}F(\omega)$ from both
meshes is plotted 
in Fig. \ref{spectralfncomp}.  For this 20 GPa case a large
broad peak dominates the 15-35 meV range, with less intense coupling 
extending on up to 70 meV.  Compared to its value from the coarser
mesh, it is much smaller in the 7-20 meV range.
The intensity in the low frequency region is clearly
from the transverse branch phonons, while the less impressive
(but still substantial) intensity in the high frequency region is
from the longitudinal phonon branch. 

The frequency resolved coupling strength
$\alpha^{2}(\omega)$, also shown in Fig. \ref{spectralfncomp} 
is dramatically changed by using the finer Q mesh.  There is now a
broad peak around 12 meV, still indicative of comparatively stronger 
coupling in the low 
frequency regime.   However, the overriding peak in the 7-16 meV 
range has disappeared, showing that is was an artifact of the 
sampling of the volume near the point K using only a 12$^3$ mesh. 
This feature is reminiscent of results calculated by Yin {\it et al.},
\cite{yin} who have found an even narrower peak in $\alpha^2(\omega)$
growing at lower frequencies (2-5 meV) in elemental Y under pressure.
This peak also was traced to incipient instability for a transverse
branch near the zone boundary.

We now concentrate on the difference in T$_c$, and the interpretation of
the distribution of the coupling, that we obtain from our denser Q mesh
of data (we note that the quantities we discuss are still not 
fully converged).  The results of this much better zone sampling are
displayed in Table \ref{table}.  Startlingly, the value of $\lambda$ 
has dropped from 3.1 to 1.0, (which was also obtained by Profeta {\it et al.}
\cite{profeta} )reflecting the fact that the
low energy region of $\alpha^2 F$ was (and still is) reliant on large
contributions (from near K and perhaps elsewhere) that vary sharply with Q.  
For the determination of T$_c$, the strong reduction in
$\lambda$ is considerably compensated 
by increases in the frequency
moments.  The logarithmic moment $\omega_{log}$ {\it increases}
by a factor of 2.7, and the second moment by 40\%. At a similar volume, Shi and
Papaconstantopoulos \cite{shi} obtain $\lambda$ = 0.5 from rigid muffin
tin calculations. One source of difference is the value $\langle\omega\rangle
\approx$ 325 K (scaled from the P = 0 Debye frequency which is our calculated 
value of $\omega_{1}$.

\begin{table}
\caption{From the calculated $\alpha^2F(\omega)$ at 20 GPa for both 
$Q$ meshes, this table provides:
the logarithmic, first, and second moments of the frequency; the value of
$\lambda$; T$_c$ (K) calculated using $\mu^*$=0.13 and $\mu^*$=0.20;
and two simple measures of $\alpha^2 F$ (see text).}
\begin{center}
\begin{tabular}{|c|c|c|c|c|c|c|c|c|}
\hline
$Q$ & $\omega_{log}$ &$\omega_1$& $\omega_2$ & $\lambda$ & T$_c$ & T$_c$ & 
 $\lambda\omega_1$ & $\sqrt{\lambda \omega_2^2}$ \\
 mesh & (K) & (K) & (K) &  & ($\mu^{*}$=0.13) & ($\mu^{*}$=0.20) & (K)&(K)\\
\hline\hline
12$^3$  &  ~81 & 115 & 176  &   3.1  & 20 & 16   & 357   & 310 \\
24$^3$  &  217 & 255 & 297  &   1.0  & 13 & ~3.4  & 250   & 294\\
\hline
\end{tabular}
\end{center}
\label{table}
\end{table}

We have obtained T$_{c}$ using the Allen-Dynes equation\cite{AD}
\begin{equation}
T_{c} = \frac{\omega_{log}}{1.2} \Lambda_1
  \Lambda_2
  exp\{- \frac{1.04(1+\lambda)}{\lambda - \mu^{*}(1 + 0.62\lambda)} \}
\end{equation}
where $\Lambda_1, \Lambda_2$ are strong coupling corrections that 
depend on $\lambda$, $\mu^*$, and 
the ratio $\omega_2/\omega_{log}$, which is a measure of the shape of $\alpha^2 F$.
These $\Lambda$ factors are important for $\lambda$=3.1 but have little effect for
$\lambda\sim$1 or less.  For the commonly used value of $\mu^*$=0.13,
our more nearly converged results give T$_c$=13 K vs. the earlier
estimate of 20 K, only a 35\% drop even though $\lambda$  decreases
by more than a factor of three.   For $\mu^*$=0.20, T$_c$ = 3.4 K.
These values should be compared with the nearly hydrostatic value
T$_c$ = 6 K at 20 GPa found by Deemyad and Schilling.\cite{schilling}

\begin{figure}[tb]
\rotatebox{-00}{\resizebox{6.5cm}{6.5cm}{\includegraphics{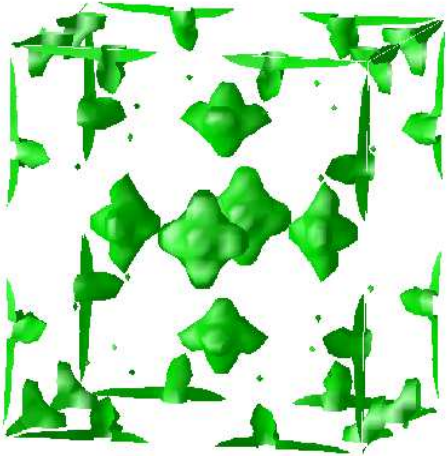}}}
\caption{(color online)
Isosurface plot of $\gamma_{Q\nu}$ = 0.054 meV for the longitudinal branch.
The box contains the $\Gamma$ point at the center and at each corner for 
this fcc structure.  The amplitude is high inside the jack-shaped region
midway between $\Gamma$ and X, corresponding to the large linewidths evident
in Fig. \ref{linewidth}. [Unfaithful interpolation at the edges of the
box account for the extra structure in those regions; the box edges are
also $\Gamma$-X-$\Gamma$ lines.]}
\label{isogamma}
\end{figure}

\section{Analysis and Summary}
This apparent weak dependence of T$_c$ on 
$\lambda$ reflects the observation,
made several times in the past,\cite{carbotte} that, for providing an
estimate of T$_c$ with averages of $\alpha^2 F$, the combination of 
$\lambda$ and some frequency moment separately is not the best choice.  This 
observation is connected with the observation that coupling at
low frequency, which strongly enhances $\lambda$, has only a weak effect
on T$_c$.  Carbotte
and collaborators\cite{carbotte} argued that the product $\lambda \omega_1$ 
(with $\omega_1$ being the first moment as defined by Allen and
Dynes\cite{AD})
is nearly proportional to T$_c$ for 1.2$\leq \lambda \leq$ 2.4; this quantity
is proportional to the area under $\alpha^2 F$ and 
corresponds to a zone integral 
of $\gamma_{Q\nu}/\omega_{Q\nu}$. 
They were not including strong coupling corrections to T$_c$,\cite{AD} however, 
which were unknown at the time of the original work. 
The strong-coupling limit\cite{AD} gives T$_c \propto$ 
$\sqrt{\lambda \omega_2^2}$; this product $\lambda \omega_2^2$ corresponds
to a zone integral of $\gamma_{Q\nu}$.  We find that, for the fine mesh
compared to the coarse (see the Table): 
$\lambda$ decreases by $70$\%; T$_c$ drops
by 35\% (at $\mu^*$=0.13); the Carbotte
product decreases by 30\%; the strong-coupling limit factor drops by
only 5\%.  Hence the Carbotte ratio follows the change in T$_c$ best, of
these choices.

The behavior of $\lambda_{Q\nu} \propto \gamma_{Q\nu}/\omega_{Q\nu}^2$ 
does not faithfully represent
the relative importance of modes for T$_c$ except in the weak-coupling limit.
In the strong coupling limit (which we do {\it not} claim is the case for
Li) it is $\gamma_{Q\nu}$ that gives the better picture.  For another
perspective we display in Fig.
\ref{isogamma} an isosurface of $\gamma_{Q\nu}$ for the ${\cal L}$
modes.  As expected from Fig. \ref{linewidth}, it is dominated by 
contributions near the center of the (100) line.  The isocontours form
a blunt jack-like shape although the true symmetry is only four-fold.
Our corresponding plots for $\lambda_{Q\nu}$ are almost like those of
Profeta {\it et al.}\cite{profeta} except that we seem to find somewhat
larger contributions near the zone center.

In this paper we have clarified and supported our earlier 
suggestion\cite{deepa} that denser
zone sampling would be necessary to obtain accurate numerical results.
We find in addition that careful sampling of the fine structure in the
coupling is required even for obtaining a correct understanding of the
microscopic origin of the pressure-induced increase in T$_c$.
In particular, we call into question the view (easily surmised from the
earlier work\cite{profeta,deepa} even if not so claimed) that the impressive
value of T$_c$ in Li under pressure
arises either from soft phonons near K, or from
a box-like set of regions in momenta along ($\eta,\eta,\zeta$) with $\eta \sim
0.7$.  These questions, and the strong polarization
dependence of matrix elements, will be addressed in further work.

\section{Acknowledgments}
We acknowledge important advice from 
S. Y. Savrasov on the linear response code, and interactions with
A. Lazicki, R. T. Scalettar, and C. S. Yoo.
We thank U. Nitzsche for assistance in the use of the IFW cluster to perform
the calculations. 
This work was supported
by National Science Foundation grant Nos. DMR-0421810 and DMR-0312261.
J.K. was supported by
DOE grant FG02-04ER46111, and H.R. was supported by DFG
(Emmy-Noether-Program).

\end{document}